\documentstyle[11pt,epsf]{article}

\def\beq{\begin{eqnarray}}
\def\eeq{\end{eqnarray}}
\def\bea{\begin{eqnarray*}}
\def\eea{\end{eqnarray*}}

\def\centeron#1#2{{\setbox0=\hbox{#1}\setbox1=\hbox{#2}\ifdim
\wd1>\wd0\kern.5\wd1\kern-.5\wd0\fi
\copy0\kern-.5\wd0\kern-.5\wd1\copy1\ifdim\wd0>\wd1
\kern.5\wd0\kern-.5\wd1\fi}}
\def\ltap{\;\centeron{\raise.35ex\hbox{$<$}}{\lower.65ex\hbox{$\sim$}}\;}
\def\gtap{\;\centeron{\raise.35ex\hbox{$>$}}{\lower.65ex\hbox{$\sim$}}\;}

\def\singleandthirdspaced{\baselineskip=\normalbaselineskip\multiply
    \baselineskip by 130\divide\baselineskip by 100}
\def\singlespaced{\baselineskip=\normalbaselineskip}

\newcommand{\newc}{\newcommand}
\newc{\qbar}{{\overline q}}
\newc{\Kahler}{K\"ahler }
\newc{\deltaGS}{\delta_{\rm GS}}
\begin{document}
\begin{titlepage}
\begin{flushright}
{\large hep-th/0203066 \\ SCIPP-2000/04\\
}
\end{flushright}

\vskip 1.2cm

\begin{center}

{\LARGE\bf The Phenomenology of Cosmological Supersymmetry
Breaking}

\vskip 1.4cm

{\large  Tom Banks$^*$}
\\
\vskip 0.4cm
{\it Santa Cruz Institute for Particle Physics,
     Santa Cruz CA 95064  } \\

\vskip 4pt

\vskip 1.5cm

\begin{abstract}
We point out phenomenological consequences of the assumption that
Supersymmetry breaking is of cosmological origin.

\end{abstract}

\end{center}

\vskip 1.0 cm

\end{titlepage}
\setcounter{footnote}{0} \setcounter{page}{2}
\setcounter{section}{0} \setcounter{subsection}{0}
\setcounter{subsubsection}{0}

\singleandthirdspaced


\section{Introduction}

The suggestion\cite{tbfolly} that Supersymmetry (SUSY) breaking is
entirely of cosmological origin has met with considerable
skepticism. To a large extent, this is due to my own failings.  I
have been unable to present either a conclusive calculation of the
large SUSY breaking effects on local physics that I claim to
follow from the existence of a cosmological horizon, nor a
definitive demonstration of the inconsistency of the old paradigm
of SUSY breaking in a Lorentz invariant vacuum state of M-theory.

The purpose of the present paper is not to engage in further
polemics on these issues, but rather to emphasize the
phenomenological consequences that follow from the assumption of
Cosmological SUSY Breaking (CSB).  I begin with a brief
recapitulation of the arguments for CSB, and then demonstrate that
they lead to dramatic conclusions.  In CSB, the cosmological
constant (in Planck units) is a (perhaps the only) tunable
parameter of the theory.  It takes on discrete values, with an
accumulation point at zero.   The basic hypothesis of the theory
is that the limiting system is a Super Poincare Invariant vacuum
state of M-theory.  I will argue that this state cannot belong to
a continuous moduli space of vacua with a boundary where the
semiclassical approximation is valid.   It must be an isolated
point.  Furthermore it must be a point of enhanced (discrete)
R-symmetry.   These considerations imply that the limiting theory
has four Poincare invariant dimensions with only minimal SUSY.
There are a number of other consequences as well, but the low
energy arguments presently available do not allow one to uniquely
predict the low energy theory.

In this paper I have adopted the following strategy.  I present
the general phenomenological implications of Cosmological SUSY
Breaking and then combine them with known facts about the low
energy theory to find constraints on the low energy description
of SUSY breaking.  One is led to a system with one or more $U(1)$
gauge groups beyond the standard model and a collection of
nonchiral standard model fields that transform under these new
gauge symmetries.  There are Fayet-Iliopoulos (FI) terms for the
new $U(1)$ fields, that are crucial to low energy SUSY breaking.
It is tempting to view these new $U(1)$ fields as the origin of a
Froggat-Nielsen \cite{fn} model of flavor physics, but I will
mostly leave this aspect of the problem to future work. In the
bulk of the paper, I freely mix up constraints from Cosmological
SUSY Breaking with strictly phenomenological constraints, but
make a careful separation between them in the Conclusions.

The point of view on SUSY breaking that I have elaborated is part
of a rather different vision of the structure of the quantum
theory of gravity than is currently held by the majority of the
community.  I have tried to avoid contaminating the present paper
with long diatribes about background independence {\it etc.}, but
in case I accidentally refer to the general framework I am writing a 
companion paper, which outlines my general viewpoint.

Finally, a note about scales.   The models that I will present in
this paper look like models of low energy SUSY breaking.  Most
previous attempts to make low energy models insisted on
explaining the TeV scale in terms of low energy gauge dynamics.
It is the essence of the idea of CSB that the origin of the low
scale of SUSY breaking cannot be understood at low energies.
Rather, the $\Lambda \rightarrow 0$ limit is viewed as a critical
phenomenon, and $m_{3/2} \sim (\Lambda / M_P^4 )^{\alpha} M_P$.
where $\alpha$ is a critical exponent determined by dynamics near
the horizon of de Sitter space. In \cite{tbfolly} I conjectured
that $\alpha = {1\over 4}$ in order to reproduce an old guess
about the relation between SUSY breaking and the TeV scale.  In
fact, given the current stage of knowledge $\alpha$ could be
anything.  Thus I cannot argue that CSB rules out Hidden Sector
SUSY breaking models.  Nor can I rule out the possibility that
different terms in the low energy effective Lagrangian might be
controlled by different critical exponents. One might for example
imagine that some of the quark mass hierarchies could be
explained by a variety of critical exponents, and would thus
depend on very high energy dynamics.

Since there are so many possibilities, I have decided to
concentrate on the simplest one, presented in \cite{tbfolly}.
That is, the low energy effective Lagrangian is a supersymmetric,
R symmetric system, with no relevant operators, and all
irrelevant operators scaled by the unification or Planck
scales.   This is modified by dimensionful terms which break both
SUSY and R symmetries (the SUSY breaking must be spontaneous, for
reasons explained below), and which are assumed to be
characterized by a single scale, near a TeV.

\section{\bf Basic framework}

The basic tenets of the CSB conjecture are simple

\begin{itemize}

\item There are no Poincare invariant, SUSY violating theories of
quantum gravity.

\item A positive cosmological constant measures the number of physical states
in the Hilbert space of quantum gravity on asymptotically deSitter
(AsdS) space.  As such it is a fundamental and freely tunable
parameter in the theory, rather than the result of a calculation
which gives the low energy effective Lagrangian.

\item Supersymmetry is of course broken in dS space.  In the limit of
vanishing cosmological constant the theory approaches a
supersymmetric theory in asymptotically flat space.  The rate at
which the gravitino mass approaches zero is (in Planck units)
$\Lambda^{\alpha}$, where the critical exponent $\alpha$ is not
calculable in low energy effective field theory.  It reflects the
dynamics of the huge degenerate spectrum of near horizon
states\footnote{In previous work, I have characterized the high
energy states as huge black holes.  In fact, most of the states
in dS space are near horizon states.} that appear in the
asymptotically flat limit.

\end{itemize}

I will not pause here to justify these statements.  The reader can
consult \cite{tbfolly} and the companion to this paper, for such
justification as exists.   My goal here is to explain how this
framework puts strong constraints on low energy phenomenology.
Let me present these constraints as an itemized list:

\begin{itemize}

\item The limit of a dS theory with vanishing cosmological constant is
an ${\cal N} = 1$, $d=4$ Super Poincare invariant and R symmetric
vacuum state of M-theory.   SUSY follows from conjecture 1 above,
and R symmetry is necessary as well, to enforce Poincare
invariance. However, the limiting theory must have a small
deformation which allows us to discuss the low energy physics of
a space with small cosmological constant.  Extended
supergravities in four dimensions, and higher dimensional
supergravities, do not have such a small deformation. Ten
dimensional IIA SUGRA has a cosmological constant but it is
quantized .  Furthermore, the spacetime it describes is the region
between BPS eight branes, and is not dS space.  Only minimal four
dimensional SUGRA (with appropriate superpotential for chiral
superfields) has dS solutions.  R symmetry (a discrete R symmetry
bigger than the $Z_2$ of $(-1)^F$ is sufficient) is necessary to
explain the vanishing of the superpotential at the supersymmetric
minimum.

\item The limiting theory has no moduli.  Here we must be a bit careful.
By moduli I mean (assuming point 1) chiral superfields with no
potential, which live in a noncompact space and obey a certain
condition at the boundaries.  To see this, consider the putative,
slightly deformed theory with small cosmological constant.  This
theory has a potential for the moduli.  If the potential goes to
zero at the boundaries of moduli space then there is a solution
of the low energy equations of motion, which describes a
spacetime that is asymptotically expanding in a subluminal
manner, and obviously has an infinite number of states.
Furthermore, even if the potential has a local dS minimum, there
is likely to be Coleman-DeLucia instanton that describes the decay of this
dS space into the asymptotically expanding one.   Thus the
quantum mechanics behind such a low energy effective theory must
have an infinite number of states and cannot be the finite model
of dS space we are studying.  Note that these conditions do not
rule out compact moduli spaces or moduli spaces that are
effectively compactified by SUSY violating potentials (for small
but nonzero $\Lambda$).  We know of no SUSic examples of such
compactified moduli spaces in string theory.

\item In the low energy effective theory, SUSY is violated
spontaneously.  This sounds like a vacuous statement.  SUSY is a
local symmetry in the field theoretic approximation, and any
explicit violation can be made to look spontaneous by introducing
a compensating goldstino field.   The nontrivial content of the
statement again comes from studying the limit of vanishing
$\Lambda$.  In this limit, SUSY is restored and the goldstino
must lie in a linear supermultiplet.  Thus, the limiting theory
must have a massless matter supermultiplet.

\item The simplest possible low energy effective theory would have
only massless chiral multiplets, $Z_i$.  The absence of mass terms
is explained by the discrete $R$ symmetry, for appropriate choices
of quantum numbers\footnote{To all orders in perturbative string
theory, masses can vanish due to a combination of stringy world
sheet symmetries, and spacetime SUSY.  In exceptional cases one
can also argue that instanton corrections are absent, and use
holomorphy to prove that certain superpotential terms vanish
exactly. However, if our chiral fields are not moduli (as implied
by the above) , no such argument applies.  We will assume that in
this case all terms in the superpotential allowed by symmetries,
are present.} .
 If $Z_i = 0$ is the point in field space corresponding to the Super
Poincare Invariant vacuum in question then the superpotential has
the form: \beq{W = A^{ijk} Z_i Z_j Z_k + o( Z^4 / M),}\eeq while
the Kahler potential is \beq {K = M^2 (\sum |Z/M|^2 + o(Z^3 /
M^3)).}\eeq  The mass scale $M$ might be the four dimensional
Planck mass.  Alternatively, we might imagine that our SUSY vacuum
incorporates Witten's mechanism \cite{witstrong} for producing a
hierarchy between the fundamental scale $M \ll M_P$ and the
Planck mass. In this case some of the fields may be scaled by
$M_P$ instead of $M$.  The constant, linear, and quadratic terms
in $W$ are not allowed by the $R$ symmetry.

In order to fine tune a large hierarchy between the gravitino mass
and the cosmological constant (something that is guaranteed at a
fundamental level by our assumption that the critical exponent
$\alpha$ is much smaller than its classical value.) we {\it must}
assume that CSB also breaks the $R$ symmetry, allowing a constant
term in $W$.  Since the $R$ symmetry is discrete, this will appear
as explicit, rather than spontaneous breaking.  The superpotential
will have corrections: \beq {\delta W = W_0 + F^i Z_i +
\mu^{ij} Z_i Z_j .}\eeq  Fine tuning of the cosmological constant
requires $F^i \sim W_0 / M_P$\footnote{I will assume that the
scale of all CSB effects is much smaller than the fundamental
scale $M$.  This amounts to assuming that $\alpha$ is not terribly
small.} . There is no obvious {\it a priori} argument for the size
of the mass matrix $\mu$. Obvious choices are $|\mu | \sim {(W_0 /
M_P)}^{1/2}$ and $|\mu | \sim W_0^{1/3}$. I have not been able to
imagine an argument which makes $\mu$ much smaller than this. For
either choice (and a wide range of other less plausible ones), the
superpotential has a SUSY minimum at a value of $Z_i$ where the
low energy effective theory is valid.  Since the superpotential is
generically nonvanishing, this is a SUSic, AdS vacuum. Thus, {\it
the purely chiral scenario is not consistent with CSB. The
hypothetical SUSY low energy theory does not admit a small
perturbation which is a dS space.} The hypothesis of CSB thus
requires that the low energy theory contain gauge fields.

\item Many gauge models are ruled out as well.  For example, pure
nonabelian SUSY gauge theories always spontaneously break $R$
symmetry.  They cannot appear in the low energy effective action
of a theory with CSB.  This is also true of a large class of
models with chiral fields transforming under a non abelian gauge
group ({\it e.g.} $SU(N_C)$ with $N_F \leq N_C$).  I have not
attempted to make a systematic delineation of the boundaries of
this class.  {\it It is notable however that the Supersymmetric
Standard Model with two or more generations does not suffer from
this problem.}  In the absence of SUSY breaking, the low energy
theory of the $N_g \geq 2$ SSM is infrared free or
superconformally invariant. It is consistent with the preservation
of a variety of discrete $R$ symmetries.

\item From now on I will concentrate on hypothetical vacua whose
low energy theory includes the SSM.  This restriction is not
derived from any low energy consistency condition.  The
explanation of why the particular groups and representations of
the SSM arise can only come from a complete understanding of
M-theory.  According to our principles, it is not permissible to
simply add soft SUSY breaking terms to the SSM Lagrangian.  The
cosmological constant can be tuned to zero, bringing the dynamics
that breaks SUSY into the low energy regime (how low depends on
the value of $\alpha$).  There are no SUSY terms that can be added
to the SSM Lagrangian to spontaneously break SUSY\footnote{Except
a Fayet-Iliopoulos (FI) term for hypercharge, which has well known
phenomenological problems.  It is also inconsistent with the idea
of grand unification, since the origin of the hypothetical FI term
comes from CSB dynamics well above the Planck scale.} .  Thus, the
hypothesis of CSB requires us to introduce further low energy
fields to explain SUSY breaking.  We will call this the SUSY
breaking sector (SSS). There are several possible scenaria for the
dynamics of the SSS.

\item There may be an interesting class of models where the SSS is
a nontrivial superconformal fixed point theory.  CSB would add
relevant perturbations to these models that spontaneously break
SUSY.  For example, consider the famous $(3,2)$ model of dynamical
SUSY breaking.  If we add  chiral fields in $(3 + \bar{3}, 2),
(1,2)$ and $(3 + \bar{3}, 1)$  to this model we can find a
nontrivial superconformal window. Mass terms for these vectorlike
fields would be forbidden by the discrete $R$ symmetry, and
induced by CSB.  For a range of values of the $SU(3)$ and $SU(2)$
couplings, the low energy dynamics will look like a superconformal
theory which crosses over to a massive theory with dynamical SUSY
breaking.  Such theories might be coupled to the SSM by marginal
({\it i.e.} realizing the standard model gauge group as a flavor
group of the superconformal theory) or irrelevant couplings.  I
know too little of the taxonomy of superconformal theories to make
a survey of this class of models in this paper.

\item Restricting our attention to theories in the vicinity of
Gaussian fixed points, there seems to be only one mechanism for
generating SUSY breaking consistent with our rules.  We must
introduce a new $U(1)$ gauge theory and assume that CSB produces
an FI term for it, as well as mass terms for chiral fields charged
under the $U(1)$.  This spontaneously breaks SUSY because of the
competition between $F$ and $D$ term constraints.  One possibility
of coupling this SSS to the SSM is through irrelevant terms. If
these terms were scaled by the fundamental scale $M$, this would
generate squark masses of order $F/M$, where $m_{3/2} \sim F/M_P
= 10^{-120\alpha} M_P$.  The experimental bounds would be
satisfied if $\alpha < 1/6$.  However, to get comparable gaugino
masses we would have to introduce singlet fields as well.
Moreover, there is a serious problem with the $\mu $ term of the
SSM.  In the SUSY limit it can be taken to vanish because of R
symmetry.   However, as noted above, the smallest reasonable
value for the induced $\mu$ term from CSB is of order $\sqrt{F}$
which is much larger than any SUSY breaking mass scale.
$SU(2)\times U(1)$ will remain unbroken and the model is
incompatible with phenomenology. The simplest model which might
avoid this disaster introduces vectorlike standard model
multiplets (to preserve coupling unification we probably want
complete $SU(5)$ multiplets) with vectorlike coupling to the new
$U(1)$.  All conventional standard model fields are taken $U(1)$
neutral. For a range of values of the CSB induced FI term, and
mass terms for these multiplets the tree level vacuum breaks SUSY
and preserves the entire standard model group.   One may hope
that the radiative corrections due to the top quark coupling will
break $SU(2) \times U(1)$ in the standard fashion.  SUSY breaking
in the standard model will occur more or less as in gauge
mediation \cite{gaugemed}, and there will be no problems with
flavor changing processes.  Note however that there are important
differences.  First of all, the $\mu$ term is induced at tree
level in the effective theory.  Secondly, SUSY breaking is not as
soft as in gauge mediation with dynamical SUSY breaking.  The
SUSY breaking terms are in the effective Lagrangian all the way
up to the GUT scale.  We will discuss this model in more detail
in the next section, and argue that one cannot in fact achieve
$SU(2) \times U(1)$ breaking by radiative corrections. This
indicates the need for more ambitious models, in which
conventional fields carry charge under the new $U(1)$. These
models are highly constrained.  We will discuss them in the next
section as well.

\end{itemize}

\section{Models}

The simplest model, allowed by the considerations of the previous
section, which breaks SUSY without breaking charge and color,
contains a number $N_5$ of ${\bf 5}$ and $\bar{\bf 5}$
representations of $SU(5)$\footnote{We presume $SU(5)$ is broken
to the standard model by a higher dimensional mechanism and that
the conventional Higgs bosons are the only incomplete four
dimensional $SU(5)$ multiplet in the low energy spectrum.}  We
denote these fields by $F^a$ and $\bar{F}_a$. We must insist that
$N_5 < 4$ to preserve perturbative coupling unification. These
multiplets carry a vectorlike representation of $U_F (1)$.  There
is also an anomaly free collection of $SU(5)$ singlet fields
charged under $U_F (1)$.

When the cosmological constant vanishes, the arguments of the
previous section tell us that we must have an R symmetric SUSic
vacuum state. We also assume that the Fayet-Iliopoulos term for
$U_F (1)$ vanishes in this limit.   If it did not, it would have
to be at least of order $M_{GUT}$, and $U_F (1)$ would not appear
in the low energy theory. We assume that mass terms $m^a_b F^a
\bar{F}_b$ as well as any $U_F (1)$ invariant mass terms for the
singlets are forbidden by the R-symmetry. The $\mu$ term for the
standard model Higgs fields is also forbidden by this R-symmetry.

When the cosmological constant is turned on, R violating mass
terms, a constant $W_0$ in the superpotential, and an FI term are
generated.  In principle the critical exponents for all of these
terms could be different and this could generate a hierarchy of
scales.   I have explained in the introduction why I choose to
restrict attention to the simplest possibility. That is, all of
these dimensionful parameters except $W_0$ are determined in
terms of a single mass scale $M_P({\Lambda \over M_P^4} )^{\alpha}
M_P$ with $\alpha \sim 1/8$ and dimensionless numbers of order
$1$.

For a range of these dimensionless coefficients, the resulting
effective potential has an absolute minimum at which $U_F (1)$
and SUSY are broken, and the standard model gauge group is
preserved.  The fields $F^a$ and $\bar{F}_b$ obtain SUSY
violating masses at tree level from the expectation value of the
$U_F (1)$ D term.

It appears however that this model will not spontaneously break
$SU(2) \times U(1)$.   The $\mu $ term induced by CSB gives a
positive mass squared to both Higgs doublets.  Radiative
corrections will give these fields SUSY violating mass terms as
well, and these can be negative.  However, they are suppressed
relative to the $\mu$ term, by a two loop factor.  Indeed, at one
loop, only standard model gauginos will get SUSY violating
masses, since the tree level SUSY breaking affects only the new
vectorlike multiplets and the singlets.  Note that $SU(3) \times
SU(2) \times U(1) \times U_F (1)$ gauge symmetry does not allow
renormalizable couplings between the chiral multiplets in the
standard model and the new fields.  Nonrenormalizable couplings
are scaled by the GUT scale and give only tiny corrections. Even
after renormalization group running, is seems unlikely that the
two loop terms can compete with the positive mass squared from the
$\mu$ term.

Thus, we are forced to give $U_F (1)$ quantum numbers to standard
model fields.  This is of course very interesting.  The simplest
possibility is to give equal and opposite $U(1)_F$ charges to the
up and down Higgs.  The $\mu$ term is still allowed but now we
can break $SU(2)\times U(1)$ at tree level by appropriate
adjustment of parameters.  This kind of model has two problems.
There are no Yukawa couplings between the Higgs and quarks and
leptons, and the electroweak breaking scale is the same as the
mass scale of a host of new particles, with standard model
couplings. It is hard to see how one could get the top quark mass
right, or avoid problems with precision electroweak
data\footnote{I would like to thank D.E. Kaplan for a discussion
of this point.}.

The latter problem seems to be shared by any model in which the
tree level $\mu$ term appears.   Since we are only allowed to
invoke local gauge symmetries, the $U(1)_F$ must forbid the $\mu$
term.   This generates an $SU(2)^2 \times U(1)_F$ anomaly, unless
we give appropriate $U(1)_F$ quantum numbers to quarks and
leptons.

We are thus driven to a class of models in which $U(1)_F$ acts as
a family symmetry \cite{annetal}.  These models are strongly
constrained.  Since I have not yet found any satisfactory models,
let me simply list the constraints.
\begin{itemize}

\item The model consists of the standard model, plus vectorlike
standard model fields (including singlets) and one or more new
$U(1)$ gauge symmetries.  Call the group of these symmetries
$U(1)_F$ even if it is a product of several $U(1)$ groups.

\item All gauge symmetries are anomaly free.

\item All relevant couplings are of the same order of magnitude.
Irrelevant operators are scaled by either the GUT or Planck
scales.

\item The $\mu$ term is forbidden by $U(1)_F$ and is replaced
by a trilinear coupling $SH_u H_d$ .  $S$ gets a small VEV from
loop corrections to generate a $\mu$ term smaller than the scale
of masses of all the new vectorlike standard model matter.  The
$\mu$ scale will also be the scale of electroweak symmetry
breaking.

\item $U(1)_F$ must commute with
$SU(5)$ or a larger GUT group. Otherwise it would have to be part
of a nonabelian group and could not get an FI D term.

\item All fields charged under the standard model, except for the
standard Higgs bosons, must come in complete GUT multiplets.

\item The vectorlike standard model matter should not lead to
Landau poles that conflict with perturbative unification. This is
a very strong constraint.

\item $U(1)_F$ must explain the quark and lepton mass and mixing
hierarchies, and provide adequate suppression of baryon and
lepton number violating operators.

\end{itemize}

It seems quite plausible that these constraints are so strong
that they have no solution at all.  I have not yet been able to
prove that.

\section{\bf Conclusions}

The reader should be convinced by now that CSB has profound
phenomenological implications.  It implies that the world is four
dimensional and is described at low energies by a small
perturbation of a model which is supersymmetric and either
infrared free or superconformally invariant in the infrared.
Within the class of Lagrangian models this restricts us to either
purely abelian models or nonabelian models with appropriate matter
content. For example, the standard model gauge group requires at
least two generations of quarks and leptons.

When perturbed by the most general set of relevant supersymmetric
perturbations, the superconformal model must spontaneously break
SUSY.  Within the class of Lagrangian models, we saw that the
only such models were those with FI D terms for some set of
$U(1)$ gauge fields.   General considerations show that these
$U(1)$'s must be external to the standard model, and in fact
commute with a GUT group.  We then showed that phenomenological
considerations (primarily a viable mechanism for electroweak
breaking) led us to a highly constrained set of models in which
the new $U(1)$'s act as family (Froggatt-Nielsen) symmetries. The
constraints on these models are very strong and there may be no
models that satisfy them.

If that is the case, there are two avenues of retreat for the
true believer in CSB.   One can consider, instead of models with
a Lagrangian description, nontrivial fixed point theories.
Spontaneous SUSY breaking by relevant perturbations of
superconformal fixed point theories has been studied
\cite{lutysundrum}, but not extensively.

Alternatively one could introduce the hypothesis that different
terms in the low energy effective Lagrangian had different
scaling behavior as the cosmological constant is taken to zero.
This could change some of our ground rules at crucial points. It
might also link some of the observed hierarchies in low energy
particle physics directly to physics at the very highest energies.
There are both pleasant and unpleasant aspects to such a linkage.
On the one hand, it would prevent us from making predictions about
the parameters that describe experiment until we understand the
full theory of quantum gravity.  On the other hand, the full
theory of quantum gravity would be directly linked to measurable
experimental quantities.  This would perhaps provide a final
answer to the hard core phenomenologist's question of why he
needs string theory.

\noindent
{\bf Acknowledgements:}

\noindent I thank A.Nelson, and D.E.Kaplan for discussions. This
work was supported in part by the U.S. Department of Energy.


\end{document}